\documentclass[twocolumn,amsmath,amssymb,prl,superscriptaddress]{revtex4-1}                    %Reference without titles
\usepackage{titlesec}
\usepackage[colorlinks,bookmarks=false,citecolor=blue,linkcolor=red,urlcolor=blue]{hyperref}
\usepackage{epsfig}
\usepackage{color}
\usepackage{subfigure}
\usepackage{graphicx}    % Include figure files
\usepackage{dcolumn}     % Align table columns on decimal point
\usepackage{lipsum}      % to move around 2-column tables !
\usepackage{braket}      % Dirac notation
\input{insbox}

\newcommand{\be}{\begin{equation}}
\newcommand{\ee}{\end{equation}}

\definecolor{drkgr}{rgb}{0.05,0.6,0.2}

\begin{document}

\title{Anisotropic Coulomb exchange as source of Kitaev\\
       and off-diagonal symmetric anisotropic couplings}

\author{Pritam Bhattacharyya}
\affiliation{Institute for Theoretical Solid State Physics, Leibniz IFW Dresden, Helmholtzstra{\ss}e~20, 01069 Dresden, Germany}

\author{Thorben Petersen}
\affiliation{Institute for Theoretical Solid State Physics, Leibniz IFW Dresden, Helmholtzstra{\ss}e~20, 01069 Dresden, Germany}

\author{Nikolay A.~Bogdanov}
\affiliation{Max Planck Institute for Solid State Research, Heisenbergstra{\ss}e~1, 70569 Stuttgart, Germany}

\author{Liviu Hozoi}
\affiliation{Institute for Theoretical Solid State Physics, Leibniz IFW Dresden, Helmholtzstra{\ss}e~20, 01069 Dresden, Germany}

\begin{abstract}
\noindent
Exchange underpins the magnetic properties of quantum matter.
In its most basic form, it occurs through the interplay of Pauli's exclusion principle and Coulomb
repulsion, being referred to as Coulomb exchange.
Pauli's exclusion principle combined with inter-atomic electron hopping additionally leads to kinetic
exchange and superexchange.
Here we disentangle the different exchange channels in anisotropic Kitaev-Heisenberg context.
By quantum chemical computations, we show that anisotropic Coulomb exchange, completely neglected
so far in the field, may be as large as (or even larger than) other contributions --- kinetic exchange
and superexchange.
This opens new perspectives onto anisotropic exchange mechanisms and sets the proper conceptual framework
for further research on tuning Kitaev-Heisenberg magnetism.
%
%
% Here we nail down the different exchange contributions to the effective magnetic couplings in a Kitaev,
% spin-orbital entangled system.
% By quantum chemical computations, we show that anisotropic Coulomb exchange defines a major interaction
% scale --- for close to ideal $j_{\mathrm{eff}}\!=\!1/2$ spin-orbital moments in $\alpha$-RuCl$_3$ under
% pressure, $\sim$45\% of the diagonal Kitaev coupling $K$ and $\sim$90\% of the off-diagonal $\Gamma'$.
% % $K$, in particular, is the source of exotic spin-orbital liquid phases.
% Anisotropic Coulomb exchange being ignored so far in Kitaev quantum magnetism research, our results
% % establish an important additional ingredient to existing Kitaev-Heisenberg exchange models.
% provide perspective onto what reliable quantitative predictions would imply: not only controlled
% approximations to deal with intersite virtual hopping but also exact Coulomb exchange.
% Interestingly, we also find a vanishingly small Heisenberg coupling $J$, which yields a $K/J$ ratio
% $\gtrsim$100 and seemingly higher chances of materializing a Kitaev quantum spin-orbital liquid.
%
\end{abstract}

\date\today
\maketitle

{\it Introduction.\,}
Magnetism has constantly been a source of new fundamental concepts in solid-state and statistical physics.
It is also important to technological applications: many devices around us are (electro)magnetic. 
Magnetism is typically illustrated through Heisenberg's textbook model of interacting atomic magnetic
moments.
Recently, however, it has become clear that for certain magnetic materials Heisenberg's isotropic
interaction picture is not applicable; their behavior can only be described through highly {\it anisotropic}
spin models.
The latter may imply completely different interaction strengths for different magnetic-moment projections
and, seemingly counterintuitive, directional dependence of the leading anisotropic coupling \cite{
Kitaev2006,Ir213_KH_jackeli_09} for symmetry-equivalent pairs of moments.
While that opens entire new perspectives in magnetism, provides the grounds for new, exotic states that
are now being revealed for the first time, and hints to potential technological applications like quantum
computation \cite{Kitaev2006}, how such anisotropies arise is not yet fully clarified: we know those may
be dominant in particular systems but do not understand in detail the underlying physics and how to tune
such interactions in the lab.

%%%%%%%%%%%%%%%%%
%%% FIGURE 1 %%%%
%%%%%%%%%%%%%%%%%
\begin{figure}[b]
\includegraphics[width=0.95\columnwidth]{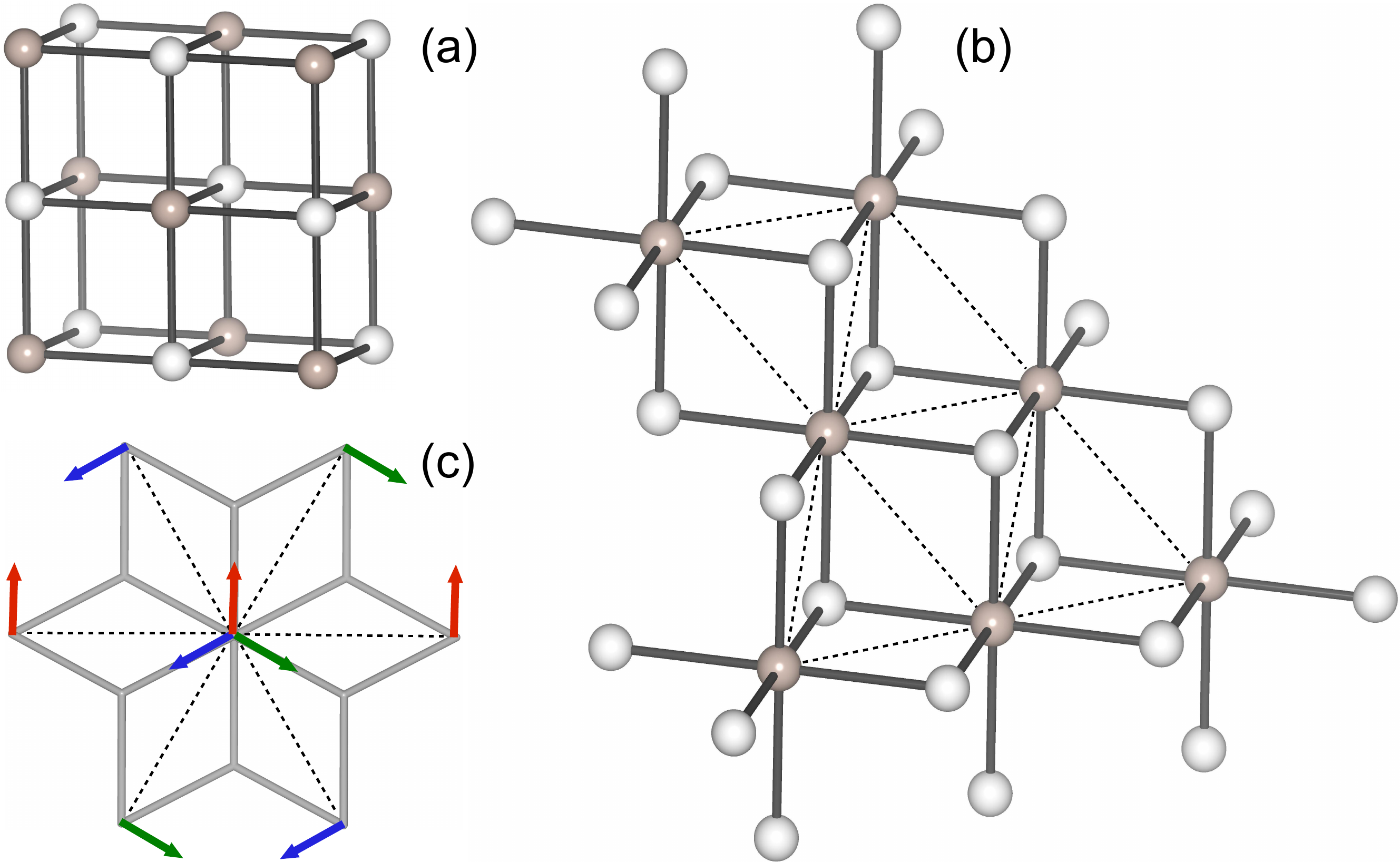}
\caption{
(a) Rocksalt-type lattice.
The M-L bonds are along either $x$, $y$, or $z$.
(b) With two different cation species ($A$, $B$) forming successive layers perpendicular to the [111]
axis, a rhombohedral $AB$L$_2$ structure is realized ---
each layer features a triangular network of edge-sharing octahedra.
Honeycomb $A_2B$L$_3$ structures are obtained when one of the cation species ($A$) occupies additional
sites within the layer of the other ($B$), corresponding to the centers of $B_6$ hexagonal rings;
in $\alpha$-RuCl$_3$, all $A$ sites are empty.
(c) On each $B_2$L$_2$ plaquette (ions not drawn), the Kitaev interaction couples only spin components
normal to the respective plaquette.
All three spin components are shown only for the central magnetic site.
}
\label{fig_1}
\end{figure}

Here we shed fresh light on the exchange mechanisms underlying symmetric anisotropic magnetic interactions,
both diagonal (i.\,e., Kitaev \cite{Kitaev2006,Ir213_KH_jackeli_09}) and off-diagonal, by using {\it ab initio}
quantum chemical computational methods.
To do so, we exploit the ladder of controlled approximations that quantum chemistry offers:
single-configuration schemes, multiconfiguration theory, and multireference configuration-interaction.
Honeycomb $\alpha$-RuCl$_3$, in particular, the relatively high-symmetry crystalline structure recently
discovered under a presure of $\approx$1.3 GPa \cite{rucl3_hp}, and triangular-lattice NaRuO$_2$ \cite{
ru112_Ortiz_2022} were chosen as benchmark Kitaev-Heisenberg material platforms.
% it displays unusually high symmetry, with only one type of Ru-Ru links, and uniform Ru-Cl-Ru bond
% angles of $\approx$93$^{\circ}$, see Fig.\;2.
%
% It turns out to be an inspired pick: our calculations show that this specific crystal structure provides
% the setting for nearly degenerate transition-metal $t_{2g}$ orbitals and close to ideal $j_{\mathrm{eff}}\!=\!1/2$
% $t_{2g}^5$ moments \cite{Ir213_KH_jackeli_09,Abragam1970}, a situation that is rarely encountered in
% solids \cite{Ir2116_grueninger_2019,Ir2116_clancy_2020,Ir2116_tsirlin_2021}.
%
We establish that a decisive contribution to the Kitaev effective coupling constant $K$ comes from
(anisotropic) Coulomb exchange, a mechanism ignored so far in the literature.
In the case of the off-diagonal ($x$-$z$/$y$-$z$) interaction $\Gamma'$, which can give rise to
spin liquid ground states by itself \cite{Ioannis_PRL}, anisotropic Coulomb exchange is even dominant,
as much as $\sim$90\% of the effective coupling parameter computed by multireference configuration-interaction.
%
% Also remarkable is the near cancellation of different exchange mechanisms for the isotropic
% Heisenberg channel, which leaves us with a fully anisotropic nearest-neighbor magnetic model.
%
% These results redefine the anisotropic pseudospin interaction frame of reference and provide
Our analysis provides unparalleled specifics as concerns $t_{2g}^5$--\,$t_{2g}^5$ Kitaev-Heisenberg
magnetic interactions and perspective onto what reliable quantitative predictions would imply: not only
controlled {\it ab initio} approximations to explicitly tackle intersite virtual excitations but also
exact Coulomb exchange; the latter is available in self-consistent-field Hartree-Fock theory, the former
in post-Hartree-Fock wave-function-based quantum chemical methods.

{\it Networks of M$_2$L$_2$ plaquettes with strong Ising-like anisotropy.\,}
Kitaev magnetism refers to anisotropic magnetic interactions $K\tilde{S}_{i}^{\gamma}\tilde{S}_{j}^{\gamma}$
that are `bond' dependent, i.e., for a given pair of adjacent 1/2 pseudospins ${\mathbf{\tilde{S}}}_{i}$
and ${\bf{\tilde{S}}}_{j}$, the easy axis defined through the index $\gamma$ can be parallel to either $x$,
$y$, or $z$ \cite{Kitaev2006}.
This can be easily visualized for layered structures of edge-sharing ML$_6$ octahedra derived from
the rocksalt crystalline arrangement (see Fig.\;\ref{fig_1}), either triangular-lattice 
$A$MO$_2$ \cite{Oles_NJP_2005}
or honeycomb $A_2$MO$_3$ \cite{Ir213_KH_jackeli_09} (and MCl$_3$) structures, where M, $A$, and L are
transition-metal, alkaline, and ligand ions, respectively:
for each of the magnetic `bonds' emerging out of a given magnetic site M, the easy axis ($x$, $y$,
or $z$) is normal to the square plaquette defined by two adjacent transition-metal ions and the two
bridging ligands.
Kitaev's honeycomb spin model has quickly become a major reference point in quantum magnetism research:
it is exactly solvable and yields a quantum spin liquid (QSL) ground state in which the spins
fractionalize into emergent Majorana quasiparticles \cite{Kitaev2006}.
The latter are neutral self-adjoint fermions that are simultaneously particle and antiparticle.
QSL ground states have been experimentally confirmed in the Kitaev-Heisenberg honeycomb systems
H$_3$LiIr$_2$O$_6$ \cite{Takagi_2018} and $\alpha$-RuCl$_3$ \cite{Science_Banerjee,Baek_2017}
% However, either in the presence of coupling-parameter disorder induced by unavoidable randomness of
% the H$^+$ cations (H$_3$LiIr$_2$O$_6$) or under external magnetic field ($\alpha$-RuCl$_3$).
but also in a triangular-lattice magnet with seemingly sizable anisotropic intersite couplings, NaRuO$_2$
\cite{ru112_Ortiz_2022}.

{\it Ru--Ru anisotropic exchange.\,}
Having QSL phases materialized on both hexagonal and triangular networks of Ru$_2$L$_2$ plaquettes makes
Ru quite special.
For insights into $t_{2g}^5$--$t_{2g}^5$ anisotropic exchange in both $\alpha$-RuCl$_3$ and NaRuO$_2$
crystallographic settings, detailed quantum chemical calculations were carried out for Ru$_2$Cl$_{10}$
and Ru$_2$O$_{10}$ magnetic units as found in the respective materials.
The adjacent in-plane RuL$_6$ octahedra coordinating those two-octahedra central units were also explicitly
included in the quantum chemical computations but using more compact atomic basis sets.
% \footnote{
% We employed relativistic pseudopotentials (ECP28MDF) and BSs (ECP28MDF-VTZ) as also used in the single-octahedron   <-- FOOTNOTE or SM ??
% computations \cite{4d_elements} for the central Ru species.
% All-electron BSs of quintuple-$\zeta$ quality were utilized for the two bridging ligands \cite{Dunning_Cl}
% and of triple-$\zeta$ quality for the remaining eight Cl anions \cite{Dunning_Cl} linked to the two octahedra
% of the reference unit.
% The four adjacent cations were represented as closed-shell Rh$^{3+}$ $t_{2g}^6$ species, using the same
% pseudopotentials (Ru ECP28MDF) and BSs (Ru ECP28MDF-VDZ [3s3p3d]) \cite{4d_elements} considered for the
% single-octahedron computations;
% the outer 16 Cl ligands associated with the four adjacent octahedra were described through minimal ANO
% BSs \cite{Pierloot1995}.}.
%
%
Complete-active-space self-consistent-field (CASSCF) optimizations \cite{olsen_bible,MCSCF_Molpro} were initially
performed with six (Ru $t_{2g}$) valence orbitals and ten electrons as active (abbreviated hereafter as
(10e,6o) active space).
% \footnote{
% The $t_{2g}$ orbitals of adjacent cations were part of the inactive orbital space.}.
%
Subsequently, two other types of wave-functions were generated, using in each case the orbitals 
obtained from the (10e,6o) CASSCF calculations\,: 
(i) single-configuration (SC) $t_{2g}^5$--$t_{2g}^5$ (i.\,e., the $t_{2g}^4$--$t_{2g}^6$ and
$t_{2g}^6$--$t_{2g}^4$ configurations which were accounted for in the initial CASSCF were excluded
in this case by imposing appropriate orbital-occupation restrictions)
% (ii) (22e,12o) complete active space configuration-interaction (CASCI) wave-functions, as full
% configuration-interaction expansions within the space defined by the Ru $t_{2g}$ and bridging-Cl
% 3$p$ orbitals,
%
and
(ii) multireference configuration-interaction (MRCI) \cite{olsen_bible,MRCI_Molpro} wave-functions having the
(10e,6o) CASSCF as kernel and additionally accounting for single and double excitations out of the
central-unit Ru $t_{2g}$ and bridging-ligand valence $p$ (either O 2$p$ or Cl 3$p$) orbitals.
By comparing data at these different levels of approximation --- SC, CASSCF, and MRCI --- it is possible
to draw conclusions on the role of various exchange mechanisms.
The CASSCF optimization was performed for the lowest nine singlet and lowest nine triplet states
associated with the (10e,6o) setting.
Those were the states for which spin-orbit couplings (SOCs) were further accounted for \cite{SOC_Molpro},
at either SC, CASSCF, or MRCI level, which yields in each case a number of 36 spin-orbit states.

%%%%%%%%%%%%%
%% TABLE 1 %%
%%%%%%%%%%%%%
\begin{table}[t]
\caption{
Nearest-neighbor magnetic couplings (meV) in high-symmetry $\alpha$-RuCl$_3$ \cite{rucl3_hp}, results
of spin-orbit calculations at various levels of theory.
% CASCI (22e,12o) stands for a full configuration-interaction within the space defined by the Ru
% $t_{2g}$ and bridging-Cl 3$p$ orbitals.
The MRCI is performed having the (10e,6o) CASSCF wave-function as kernel.
}
\begin{tabular}{l c c c c}
\hline
\hline
\\[-0.25cm]
                 &{\it{K}}  &{\it{J}}  &$\Gamma_{xy}\!\equiv\!\Gamma$
                                                      &$\Gamma_{yz}\!=\!\Gamma_{zx}\!\equiv\!\Gamma'$\\

\hline
\\
[-0.22cm]
SC               &--1.75    &0.35      &--0.11        &0.42   \\%[0.11cm]
CASSCF (10e,6o)  &--1.73    &--1.04    &0.89          &0.46   \\
%CASCI (22e,12o) &--1.72    &--1.04    &0.89          &0.47   \\%[0.11cm]
MRCI             &--3.73    &--0.03    &1.62          &0.45   \\
\hline
\hline
\end{tabular}
\label{tab:couplings}
\end{table}

% Only one type of Ru-Ru link is present in $\alpha$-RuCl$_3$ at $p\!=\!1.26$ GPa.
A unit of two nearest-neighbor octahedra exhibits $C_{2h}$ point-group symmetry, in both $\alpha$-RuCl$_3$
\cite{rucl3_hp} and NaRuO$_2$ \cite{ru112_Ortiz_2022}, implying a generalized bilinear effective spin
Hamiltonian of the following form for a pair of adjacent 1/2-pseudospins ${\bf{\tilde{S}}}_i$ and
${\bf{\tilde{S}}}_j$\,:
\begin{equation}
\label{eqn:Hamil}
\mathcal{H}_{ij}^{(\gamma)} = J{\bf{\tilde{S}}}_i\cdot {\bf{\tilde{S}}}_j + \\
                              K\tilde{S}_i^{\gamma}\tilde{S}_j^{\gamma} + \\
                              \sum_{\alpha\neq\beta} \Gamma_{\alpha\beta} \\
                              (\tilde{S}_i^{\alpha}\tilde{S}_j^{\beta} +  \\
                              \tilde{S}_i^{\beta}\tilde{S}_j^{\alpha}).
\end{equation}
The $\Gamma_{\alpha\beta}$ coefficients denote the off-diagonal components of the 3$\times$3
symmetric-anisotropy exchange tensor, with $\alpha,\beta,\gamma\!\in\!\{x,y,z\}$. 
% An antisymmetric Dzyaloshinskii-Moriya coupling is not allowed, given the inversion center.
%
The lowest four spin-orbit eigenstates from the {\it ab initio} quantum chemical output (eigenvalues
lower by $\sim$0.2 eV with respect to the eigenvalues of higher-lying excited states) are mapped for
each different set of calculations onto the eigenvectors of the effective spin Hamiltonian (1),
following the procedure described in refs.~\cite{Bogdanov_et_al,Yadav2016}\,:
those four expectation values and the matrix elements of the Zeeman Hamiltonian in the basis of the
four lowest-energy spin-orbit eigenvectors are put in direct correspondence with the respective
eigenvalues and matrix elements of (\ref{eqn:Hamil}).
%
% Having two of the states in the same irreducible representation of the $C_{2h}$ point group, such
% one-to-one mapping translates into two possible sets of effective magnetic couplings.
% The relevant array is chosen as the one whose $g$ factors fit the $g$ factors corresponding to
% a single RuCl$_6$ $t_{2g}^5$ octahedron.
%
%% We used the standard coordinate frame usually employed in the literature, different from the rotated
%% frame employed in earlier quantum chemical studies \cite{Yadav2016,Nishimoto2016,Ravi_CS} that affects
%% the sign of $\Gamma$ (see also discussion in \cite{NaRuO2}).

%%%%%%%%%%%%%%%%%
%%% FIGURE 2 %%%%
%%%%%%%%%%%%%%%%%
\begin{figure}[b]
\includegraphics[width=0.90\columnwidth]{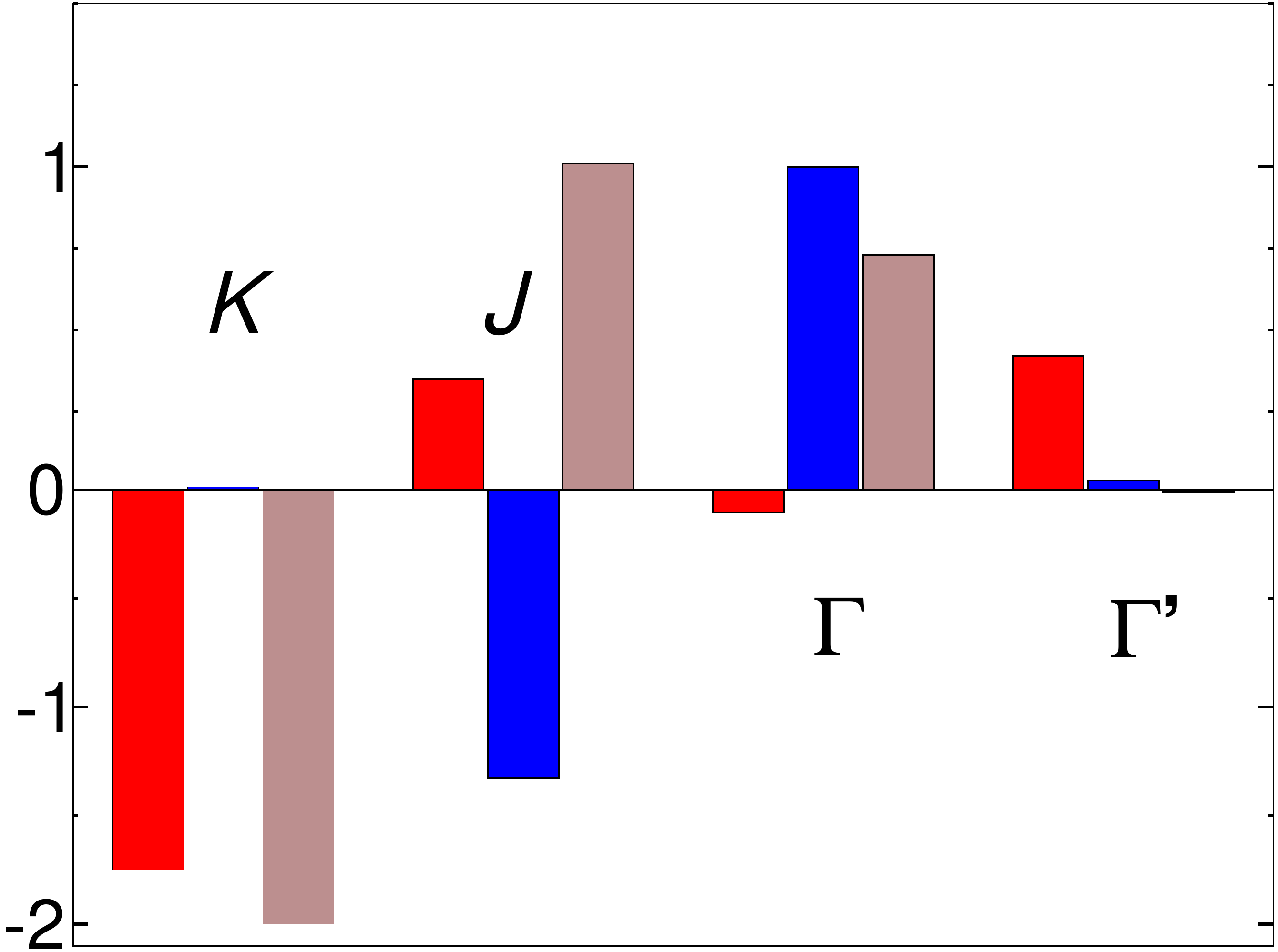}
\caption{
Exchange contributions to the intersite effective magnetic couplings in high-symmetry $\alpha$-RuCl$_3$
\cite{rucl3_hp}:
Coulomb exchange (SC results, in red), Ru($t_{2g}$)--Ru($t_{2g}$) kinetic exchange (as the difference between
CASSCF and SC data, in blue), plus contributions related to Ru-Cl-Ru superexchange \cite{Anderson_1959,
KANAMORI1959,Khaliullin_PRL_2010,J3_winter_16} and so called dynamical correlation effects \cite{olsen_bible}
(as the difference between MRCI and CASSCF, in brown).
}
\label{contribs_13}
\end{figure}

Nearest-neighbor effective magnetic couplings as obtained at three different levels of theory (SOC
included) for $\alpha$-RuCl$_3$ are listed in Table \ref{tab:couplings}.
A very interesting finding is the vanishingly small $J$ value in the spin-orbit MRCI computations, which
yields a fully anisotropic $K$-$\Gamma$-$\Gamma'$ effective magnetic model for the nearest-neighbor
magnetic interactions and in principle increases the chances of realizing a QSL ground state.
This particular aspect will be discussed elsewhere.

Even more remarkable are the large anisotropic Coulomb exchange contributions obtained by SC calculations
with SOC.
% the SC value represents $\sim$45\% of the MRCI Kitaev coupling $K$ and as much as $\sim$90\% of the MRCI
% off-diagonal $\Gamma'$ effective parameter.
% % Anisotropic Coulomb exchange being ignored so far in Kitaev quantum magnetism research, our results
% %% establish an important additional ingredient to existing Kitaev-Heisenberg exchange models.
%
The diagonal Kitaev coupling $K$, for example, is basically the same at the lowest two levels of
approximation (first column in Table~\ref{tab:couplings}),
SC (only the $t_{2g}^5$--$t_{2g}^5$ electron configuration considered) and CASSCF (10e,6o) ($t_{2g}^5
$--$t_{2g}^5$, $t_{2g}^4$--$t_{2g}^6$, and $t_{2g}^6$--$t_{2g}^4$ configurations treated on the same
footing, where the latter type of states bring kinetic Ru($t_{2g}$)--Ru($t_{2g}$) exchange).
This indicates that intersite Ru\,$t_{2g}$\,$\rightarrow$\,Ru\,$t_{2g}$ excitations (i.e., kinetic
exchange) do not really affect $K$.
What matters as concerns the size of the Kitaev coupling are 
(i) Coulomb exchange, with a contribution of --1.75 meV, and
(ii) excitations to higher-lying states and so called dynamical correlation effects \cite{olsen_bible}
accounted for in MRCI, with a contribution of --2 meV.
Part of (ii) corresponds to Ru-Cl-Ru superexchange \cite{Anderson_1959,KANAMORI1959}, involving also
the Ru 4$d$ $e_g$ levels \cite{Ir213_KH_jackeli_09,Khaliullin_PRL_2010,J3_winter_16}.
Especially striking is the diagnosis carried out for the off-diagonal $\Gamma'$ effective interaction
parameter:
out of a spin-orbit MRCI value of 0.45 meV, 0.42 corresponds to anisotropic Coulomb exchange.
A pictorial representation of the various contributions to $K$, $J$, $\Gamma$, and $\Gamma'$ in
high-symmetry $\alpha$-RuCl$_3$ \cite{rucl3_hp} is provided in Fig.\;\ref{contribs_13}.

\begin{figure}[t]
\includegraphics[width=0.72\columnwidth]{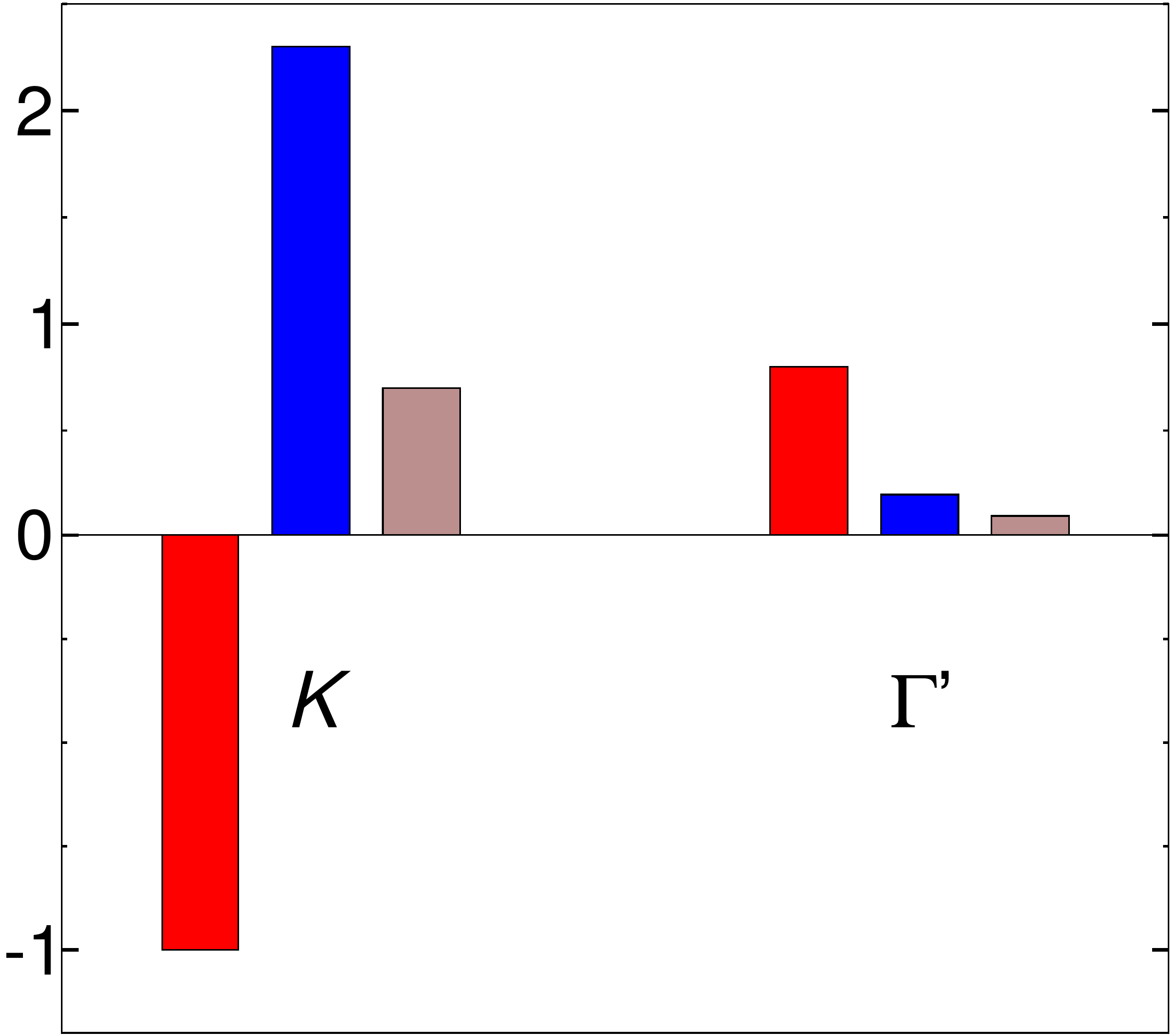}
\caption{
Exchange contributions to $K$ and $\Gamma'$ in NaRuO$_2$ \cite{ru112_Ortiz_2022}:
Coulomb exchange (SC results, in red), Ru($t_{2g}$)--Ru($t_{2g}$) kinetic exchange (as the difference
between CASSCF and SC data, in blue), plus contributions related to Ru-O-Ru superexchange and dynamical
correlations (as the difference between MRCI and CASSCF, in brown).
}
\label{contribs_112}
\end{figure}

MRCI+SOC computations for adjacent edge-sharing RuO$_6$ octahedra in triangular-lattice NaRuO$_2$
indicate that the largest nearest-neighbor coupling parameter is the isotropic Heisenberg $J$, --5.2 meV;
the other effective interactions, $K$, $\Gamma$, and $\Gamma'$, amount to 2, 3.6, and 1.1 meV, respectively,
by spin-orbit MRCI.
For better visualization, since the most important anisotropic Coulomb exchange contributions arise also
in this system for $K$ and $\Gamma'$, we depict in Fig.\;\ref{contribs_112} only the structure of these two magnetic
couplings and omit the $J$ and $\Gamma$ effective interactions, which have significantly larger absolute
values.
Plots for the latter are provided in Supplemental Material (SM).
It is seen that anisotropic Coulomb exchange represents the second largest contribution to the Kitaev $K$
and the leading underlying mechanism in the case of $\Gamma'$.

%%%%%%%%%%%%%%%%%
%%% FIGURE 3 %%%%
%%%%%%%%%%%%%%%%%

Anisotropic Coulomb exchange as found in the SC calculation (also referred to as direct exchange
\cite{Anderson_1959}) represents genuine new physics, not addressed so far in the literature.
Finding that up to $\sim$45\% of the Kitaev effective coupling constant $K$ has to do with Coulomb exchange 
and that the off-diagonal anisotropic coupling $\Gamma'$, which may give rise to spin-liquid ground
states by itself \cite{Ioannis_PRL}, comes more than 90\% from Coulomb exchange (last column in Table
\ref{tab:couplings}) obviously challenges present views and notions in Kitaev-Heisenberg quantum
magnetism research.
This is just another example illustrating the need for {\it ab initio} quantum chemical methods in order
to achieve even a qualitatively correct picture of the essential underlying physics.
Recent quantum chemical results that lead to the same conclusion refer to the role of fluctuations involving
the third and fourth electronic shells in renormalizing antiferromagnetic interactions in copper oxide compounds
\cite{Bogdanov2022}. 

To provide additional reference points, we computed the isotropic Coulomb exchange integrals
(i.\,e., without accounting for SOC) for different distributions of the Ru $t_{2g}$ holes in the
$t_{2g}^5$--$t_{2g}^5$ arrangement.
For holes in plaquette-plane 4$d$ orbitals having overlapping lobes along the Ru-Ru axis (i.e., for
$xy$-like $t_{2g}$ functions), for example, the Coulomb exchange integral amounts to --25.4 meV.
For comparison, in 3$d^9$ copper oxide compounds with corner-sharing ligand octahedra, the Coulomb exchange 
matrix element is in the region of --10 meV (from SC $d_{x^2-y^2}^1$--\,$d_{x^2-y^2}^1$ calculations)
\cite{Martin_1993,VANOOSTEN1996}.
Being aware of experimental estimates of 100--150 meV for the Heisenberg $J$, an isotropic Coulomb exchange
contribution of --10 meV can be neglected in layered cuprates.
Yet, this does not seem to be the case for $t_{2g}^5$ edge-sharing octahedra. 

The full 3$\times$3 matrix for adjacent Ru $t_{2g}^5$ sites ($i$, $j$) in high-symmetry RuCl$_3$ is 
provided in Table\;\ref{tab:directex}.
%
%% How exactly SOC and Coulomb interactions mingle to yield large {\it anisotropic} direct exchange 
%% integrals will be discussed elsewhere.
How exactly SOC and Coulomb interactions commix to yield large {\it anisotropic} Coulomb exchange integrals
remains to be analyzed in detail in future work.
The important point however is that, at the $t_{2g}^5$--$t_{2g}^5$ SC level, there is a Coulomb exchange
matrix element for each possible pair of holes --- $d_{xy}$-$d_{xy}$, $d_{xy}$-$d_{yz}$ etc.
SOC mixes up those different Slater determinants, and the resulting spin-orbit wave-functions are not
spin eigenstates.
The spin-orbit level structure in the two-octahedra problem can be reduced to an effective pseudospin
model only by introducing anisotropic Coulomb exchange matrix elements (i.\,e., the SC values provided
in Table\;\ref{tab:couplings}).

%%%%%%%%%%%%%
%% TABLE 2 %%
%%%%%%%%%%%%%
\begin{table}[t]
\caption{
% Coulomb (direct) exchange matrix elements (in meV) for each possible pair of holes.
Coulomb exchange (in meV) for Ru $t_{2g}$ holes in RuCl$_3$ $t_{2g}^5$--$t_{2g}^5$ arrangement,
obtained in each case as the difference between triplet ($d_{i,\alpha}^{\uparrow}d_{j,\beta}^{\uparrow}$,
$\alpha$,\,$\beta$\,$\in$\,\{$xy$,\,$yz$,\,$zx$\}) and singlet ($d_{i,\alpha}^{\uparrow}d_{j,\beta}^{\downarrow}$)
energies.
}
\begin{tabular}{c c c c}
\hline
\hline
\\[-0.25cm]
           \hspace{1cm}    &{\it{d}$_{xy}$} \hspace{1cm} &{\it{d}$_{yz}$} \hspace{1cm}  &{\it{d}$_{zx}$} \\

\hline
\\
[-0.22cm]
{\it{d}$_{xy}$} \hspace{1cm} &--25.4  \hspace{1cm}    &--4.7   \hspace{1cm}    &--4.7          \\%[0.11cm]
{\it{d}$_{yz}$} \hspace{1cm} &--4.7   \hspace{1cm}    &--0.6   \hspace{1cm}    &--7.0          \\
{\it{d}$_{zx}$} \hspace{1cm} &--4.7   \hspace{1cm}    &--7.0   \hspace{1cm}    &--0.6          \\
\hline
\hline
\end{tabular}
\label{tab:directex}
\end{table}

{\it Conclusions.\,}
To summarize, we resolve the exchange mechanisms giving rise to anisotropic magnetic interactions in
hexagonal and triangular networks of edge-sharing RuL$_6$ $t_{2g}^5$ octahedra. 
Different from present assumptions and exchange models relying exclusively on inter-atomic hopping 
processes, the quantum chemical analysis indicates major Coulomb-exchange contributions, to both 
$K$ and $\Gamma'$.
These findings redefine the conceptual frame within which Kitaev magnetism should be addressed.
Also, in light of the {\it ab initio} quantum chemical data, various estimates, interpretations, and
predictions based only on kinetic-exchange and superexchange models might require reevaluation --- 
what is represented in red color in Figs.\;2 and 3 is simply ignored in existing effective-model 
theories and studies.

 \ 

{\it Acknowledgments.\,}
P.\,B., T.\,P., and L.\,H. acknowledge financial support from the German Research Foundation (Deutsche
Forschungsgemeinschaft, DFG), projects 441216021 and 468093414, and technical assistance from U.~Nitzsche.
We thank J.\,Geck, G.~Khaliullin, S.\,Winter, O.\,Janson, S.\,D.\,Wilson, and U.\,K.\,R{\"o}{\ss}ler
for discussions.

\bibliography{refs_jun29}

\end{document}

\title{Supplemental Material --- Anisotropic Coulomb exchange as source of Kitaev\\
       and off-diagonal symmetric anisotropic couplings}

\author{Pritam Bhattacharyya}
\affiliation{Institute for Theoretical Solid State Physics, Leibniz IFW Dresden, Helmholtzstra{\ss}e~20, 01069 Dresden, Germany}

\author{Thorben Petersen}
\affiliation{Institute for Theoretical Solid State Physics, Leibniz IFW Dresden, Helmholtzstra{\ss}e~20, 01069 Dresden, Germany}

\author{Nikolay A.~Bogdanov}
\affiliation{Max Planck Institute for Solid State Research, Heisenbergstra{\ss}e~1, 70569 Stuttgart, Germany}

\author{Liviu Hozoi}
\affiliation{Institute for Theoretical Solid State Physics, Leibniz IFW Dresden, Helmholtzstra{\ss}e~20, 01069 Dresden, Germany}

\date\today
\maketitle

Different from the vast majority of Kitaev-Heisenberg honeycomb systems, only one type of Ru-Ru link
is present in both $\alpha$-RuCl$_3$ at $p\!=\!1.26$ GPa \cite{rucl3_hp} and NaRuO$_2$ \cite{
ru112_Ortiz_2022}.
Since a unit of two nearest-neighbor octahedra exhibits $C_{2h}$ point-group symmetry, the relevant
bilinear effective spin Hamiltonian for a pair of adjacent 1/2-pseudospins ${\bf{\tilde{S}}}_i$ and
${\bf{\tilde{S}}}_j$ reads
\begin{equation}
\label{eqn:Hamil}
\mathcal{H}_{ij}^{(\gamma)} = J{\bf{\tilde{S}}}_i\cdot {\bf{\tilde{S}}}_j + \\
                              K\tilde{S}_i^{\gamma}\tilde{S}_j^{\gamma} + \\
                              \sum_{\alpha\neq\beta} \Gamma_{\alpha\beta} \\
                              (\tilde{S}_i^{\alpha}\tilde{S}_j^{\beta} +  \\
                              \tilde{S}_i^{\beta}\tilde{S}_j^{\alpha}).
\end{equation}
The $\Gamma_{\alpha\beta}$ coefficients denote the off-diagonal components of the 3$\times$3
symmetric-anisotropy exchange tensor, with $\alpha,\beta,\gamma\!\in\!\{x,y,z\}$. 
An antisymmetric Dzyaloshinskii-Moriya coupling is not allowed, given the inversion center.

The lowest four expectation values obtained from the quantum chemical calculations (either SC, CASSCF,
or MRCI level) for both compounds (NaRuO$_2$ and high-symmetry $\alpha$-RuCl$_3$) 
and the matrix elements of the Zeeman Hamiltonian in the basis of the four lowest-energy
spin-orbit eigenvectors are put in direct correspondence with the respective eigenvalues and matrix
elements of (\ref{eqn:Hamil}).
Having two of the states in the same irreducible representation of the $C_{2h}$ point group, such
one-to-one mapping translates into two possible sets of effective magnetic couplings.
The relevant array is chosen as the one whose $g$ factors fit the $g$ factors corresponding to
a single RuCl$_6$ (RuO$_6$) $t_{2g}^5$ octahedron.

We used the standard coordinate frame usually employed in the literature, different from the rotated
frame employed in earlier quantum chemical studies \cite{Yadav2016,Nishimoto2016,Ravi_CS} that affects
the sign of $\Gamma$ (see also footnote [48] in  Ref.~\cite{rucl3_janssen_17}).
The SC label in Table I in the main text indicates a CASCI in which intersite excitations are not 
considered. This is also referred to as occupation-restricted multiple active space (ORMAS) 
scheme \cite{ormas_ivanic_03}.

We employed relativistic pseudopotentials (ECP28MDF) and BSs (ECP28MDF-VTZ) \cite{4d_elements} for the 
central Ru species.
All-electron BSs of quintuple-$\zeta$ quality were utilized for the two bridging ligands \cite{Dunning_Cl}
and of triple-$\zeta$ quality for the remaining eight Cl anions \cite{Dunning_Cl} linked to the two octahedra
of the reference unit.
The four adjacent cations were represented as closed-shell Rh$^{3+}$ $t_{2g}^6$ species, using the same
pseudopotentials (Ru ECP28MDF) and BSs (Ru ECP28MDF-VDZ [3s3p3d]) \cite{4d_elements} considered for the
single-octahedron computations;
the outer 16 Cl ligands associated with the four adjacent octahedra were described through minimal ANO
BSs \cite{Pierloot1995}.

%%%%%%%%%%%%%%%%%
%%% FIGURE S1 %%%%
%%%%%%%%%%%%%%%%%
\begin{figure}[]
\includegraphics[width=0.95\columnwidth]{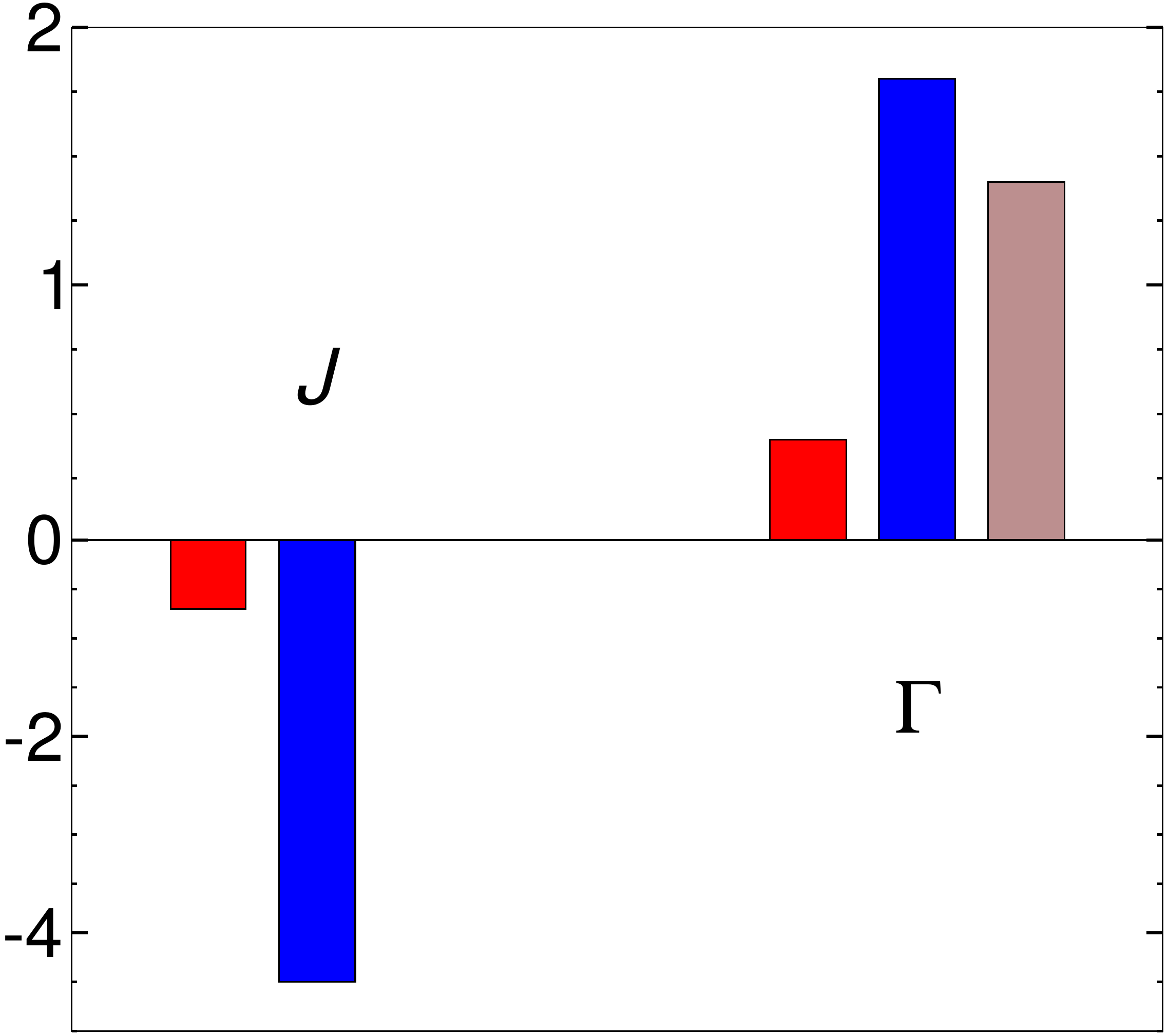}
\caption{
Exchange contributions to $J$ and $\Gamma$ in NaRuO$_2$ \cite{ru112_Ortiz_2022}:
Coulomb exchange (SC results, in red), Ru($t_{2g}$)--Ru($t_{2g}$) kinetic exchange (as the difference
between CASSCF and SC data, in blue), plus contributions related to Ru-O-Ru superexchange and dynamical
correlations (as the difference between MRCI and CASSCF, in brown);
the brown contribution of J is zero.
}
\label{NaRuO2_J_G}
\end{figure}

To derive the inter-site effective magnetic couplings of NaRuO$_2$,
clusters with two edge-sharing RuO$_6$ octahedra in the central region were considered.
The eight in-plane octahedra directly linked to the two-octahedra central unit and 22 nearby Na ions
were also explicitly included in the quantum chemical computations but using much more compact basis
sets.
We utilized energy-consistent relativistic pseudopotentials (ECP28MDF) and Gaussian-type valence basis
sets of effective quadruple-$\zeta$ quality (ECP28MDF-VTZ) for the central Ru species \cite{4d_elements}.
All-electron basis sets of quintuple-$\zeta$ quality were employed for the two bridging ligands
\cite{O-ligands} while all-electron basis sets of triple-$\zeta$ quality were applied for the remaining
eight O anions \cite{O-ligands} associated with the two octahedra of the reference unit.
The eight adjacent transition-metal (TM) cations were represented as closed-shell Rh$^{3+}$ $t_{2g}^6$
species, using non-relativistic pseudopotentials (Ru ECP29MHF) and (Ru ECP28MHF) (8s7p6d)/[4s3p2d] basis
sets \cite{Andrae1990,Martin_et_al};
the outer 22 O ligands associated with the eight adjacent octahedra were described through minimal
all-electron atomic-natural-orbital (ANO) basis sets \cite{Pierloot1995}. 
Large-core pseudopotentials were considered for the 22 Na nearby cations \cite{Fuentealba_Na}.

%A phase diagram based on the MRCI nearest-neighbor effective couplings and variable second- ($J_2$)
%and third-neighbor ($J_3$) Heisenberg interactions is shown in Fig.~\ref{fig_J2_J3}.
%The QSL2 phase is similar to the Kitaev QSL (see discussion in the main text).
%The QSL3 region denotes a frustration-induced spin liquid.
%The spin-spin correlations are small, reminiscent in some sense of a paramagnetic state.
%Since $J_2$ is expected to be tiny (see discussion in the main text), the QSL2 state can be realized
%for a wide range of $J_3$ values.

\bibliography{refs_jun29}